\documentclass{aa}  

\usepackage{graphicx}
\usepackage[varg]{txfonts}
\usepackage{xcolor}

\usepackage{xpatch}
\usepackage{hyperref}
\hypersetup{
	colorlinks=true,
	breaklinks=true,
	citecolor=blue,
	allcolors=blue,
	frenchlinks=true
}

\makeatletter
\xpatchcmd\NAT@citex
{%
	\@citea\NAT@hyper@{%
		\NAT@nmfmt{\NAT@nm}%
		\hyper@natlinkbreak{\NAT@aysep\NAT@spacechar}{\@citeb\@extra@b@citeb}%
		\NAT@date
	}%
}
{%
	\@citea
	\NAT@nmfmt{\NAT@nm}%
	\NAT@aysep\NAT@spacechar
	\NAT@hyper@{\NAT@date}%
}
{}{}
\xpatchcmd\NAT@citex
{%
	\@citea\NAT@hyper@{%
		\NAT@nmfmt{\NAT@nm}%
		\hyper@natlinkbreak{\NAT@spacechar\NAT@@open\if*#1*\else#1\NAT@spacechar\fi}%
		{\@citeb\@extra@b@citeb}%
		\NAT@date
	}%
}
{
	\@citea
	\NAT@nmfmt{\NAT@nm}%
	\NAT@spacechar\NAT@@open\if*#1*\else#1\NAT@spacechar\fi
	\NAT@hyper@{\NAT@date}%
}
{}{}
\makeatother

\makeatletter
\renewcommand*\aa@pageof{, page \thepage{} of \pageref*{LastPage}}
\makeatother

\usepackage{natbib}
\bibpunct{(}{)}{;}{a}{}{,}

\newcommand{\bcdot}{\boldsymbol{\cdot}}

\begin{document}

   \title{Why are thermally- and cosmic ray-driven galactic winds fundamentally different?}

   \titlerunning{Thermally vs cosmic-ray-driven galactic winds}

   \author{T. Thomas\inst{1}\fnmsep\thanks{\email{tthomas@aip.de}},
          C. Pfrommer\inst{1}
          and
          R. Pakmor\inst{2}
          }
   \institute{Leibniz Institute for Astrophysics Potsdam (AIP), An der Sternwarte 16, 14482 Potsdam, Germany
         \and
             Max Planck Institute for Astrophysics, Karl-Schwarzschild-Strasse 1, 85740 Garching, Germany}

   \date{Received ...; accepted ...}

  \abstract
   { 
    Galactic outflows influence the evolution of galaxies not only by expelling gas from their disks but also by injecting energy into the circumgalactic medium (CGM). This alters or even prevents the inflow of fresh gas onto the disk and thus reduces the star formation rate. Supernovae (SNe) are the engines of galactic winds as they release thermal and kinetic energy into the interstellar medium (ISM). Cosmic rays (CRs) are accelerated at the shocks of SN remnants and only constitute a small fraction of the overall SN energy budget. However, their long live-times allow them to act far away from the original injection site and thereby to participate in the galactic wind launching process. Using high-resolution simulations of an isolated Milky Way-type galaxy with the moving-mesh code \textsc{Arepo} and the new multi-phase ISM model \textsc{Crisp} (Cosmic Rays and InterStellar Physics), we investigate how SNe and CRs launch galactic outflows and how the inclusion of CR-mediated feedback boosts the energy and mass entrained in the galactic wind. We find that the majority of thermal SN energy and momentum is used for stirring turbulence either directly or indirectly by causing fountain flows, thereby self-regulating the ISM and not for efficiently driving outflows to large heights. A simulation without CRs only launches a weak galactic outflow at uniformly high temperatures and low densities by means of the thermal pressure gradient. By contrast, most of the CR energy accelerated at SN remnants ($\sim80\%$) escapes the ISM and moves into the CGM. In the inner CGM, CRs dominate the overall pressure and are able to accelerate a large mass fraction in a galactic wind. This wind is turbulent and multi-phase with cold cloudlets embedded in dilute gas at intermediate temperatures ($\sim10^5$~K) and the CGM shows enhanced \ion{O}{vi} and \ion{C}{iv} absorption in comparison to a simulation without CRs. }


   \maketitle


\section{Introduction}

Galactic winds are a key phenomenon regulating the baryon mass budget of galaxies \citep{2005Veilleux, 2018Zhang}. Their dynamical impact reaches beyond simply propelling gas from within the ISM of a galaxy into its surrounding CGM but the energy entrained with these galactic outflows can mechanically divert gas that is falling onto the galaxy away from its original trajectory. It can also heat the CGM to establish an atmosphere around the galaxy, altering the accretion of fresh gas onto the galactic disk. The main consequence of galactic winds is thus the reduction of gas in the galactic disk that would be available for subsequent star formation. 

Beyond this regulatory aspect, galactic winds are proposed to be a source of metals and magnetic fields that pervade the CGM \citep{2017Tumlinson, 2020Pakmor, 2021vandeVoort,2024Heesen}. Enriching the CGM with metals further strengths its ability to efficiently cool radiatively via atomic line transitions, adding to the complexity of this medium. Furthermore, the presence of metals in the CGM allows us to observe this medium by means of the numerous transitions of metal ions either in absorption or emission. In such observations, the CGM reveals itself as a diverse multiphase medium filled with molecular \citep{2015Leroy,2020Diteodoro}, atomic \citep{2018Martini, 2020Das} and ionized gas \citep{2014Werk, 2016Schroetter, 2020Lehner, 2023Guo}. Understanding the mechanisms driving galactic winds is essential for interpreting these observations and for elucidating their impact on the formation and evolution of galaxies. 

Stellar feedback, in the form of mechanically driven SNe or stellar winds, is mainly responsible for maintaining a self-regulated dynamical state of the ISM \citep{2022Ostriker}. Multiple massive stars born in the same giant molecular cloud (GMC) eventually explode as SNe, occurring in close proximity to one another, which leads to the formation of super-bubbles through a clustered feedback event \citep{2016GirichidisII, 2020Fielding, 2023Barnes, 2023Mayya}. Strong radiation fields originating from young and massive stars heat their environments through photoionization by extreme ultraviolet photons in HII regions or through the dust-mediated photoelectric effect caused by far ultravoilet (FUV) photons \citep{1939Stromgren, 1978Draine}. Radiative cooling  counteracts these heating processes and establishes a dynamic, multiphase ISM \citep{1969Field, 1995Wolfire}. With the advent of modern computational astrophysics, the \textsc{SILCC} and \textsc{TIGRESS} simulation campaigns were able to model this  physics-rich environment at sufficient resolution and found that the interplay of these various physical processes are in fact self-regulating the ISM and preventing it from gravitational collapse \citep{2015Walch,2023Rathjen,2017Kim,2023KimII}. 

Some of the mechanical energy of an SN shock will be used to accelerated thermal particles to mildy- and ultrarelativistic energies, giving birth to a population of high-energy CRs \citep{1987Blandford}. While they are comparatively few in number, their large individual particle energies place the pressure provided by the CR population favorable amongst other dynamical relevant pressures such as the thermal, magnetic or turbulent pressures \citep{1990Boulares, 2017Naab}. After leaving the ISM, CRs build up a reservoir throughout the CGM where their pressure affects the dynamics of infalling gas \citep{2016Salem, 2020Buck, 2021Ji}. Their passage from the ISM into the CGM as well as their redistribution in these media, is dictated by the various CR transport phenomena. CRs with energies of around a few GeV interact with their surrounding gas through plasma physical processes which scatter the CRs through their interactions with electromagnetic fields. Small-scale perturbations of the electromagnetic fields responsible for such scatterings materialize in the form CR-self-excited Alfv\'en waves through the streaming instability \citep{1969Kulsrud, 1975SkillingI, 2013Zweibel} or the non-resonant Bell instability \citep{2004Bell}, while the intermediate-scale instability drives whistler waves unstable \citep{2021Shalaby, 2023Shalaby}. CRs can furthermore be scattered at Alfv\'en waves driven unstable by the dust-mediated streaming instability \citep{2021Squire} or at micro-mirrors \citep{2023Reichherzer}. Intermittent large-scale high-amplitude structures of the magnetic field are also proposed as potential scattering entities \citep{2023Kempski, 2023Lemoine}.

CRs are of particular interest as the momentum they impart on their surrounding media can drastically alter is dynamics \citep[for recent reviews, see][]{2017Zweibel, 2023Ruszkowski, 2023Owen}. In the context of galactic winds, it has been hypothesed based on theoretical arguments that CRs might be an important agent for driving galactic winds \citep{1975Ipavich, 1991Breitschwerdt}. 
This conclusion has been supported by various numerical simulations of galaxy formation with different simulation approaches such as tallbox simulations targeting small patches of the ISM at high resolution \citep{2016Girichidis, 2018Farber, 2019Holguin, 2021Armillotta, 2023Simpson}, simulations of isolated galaxies \citep{2012Uhlig, 2014Salem, 2016PakmorIII, 2017Ruszkowski, 2020Dashyan, 2022Farcy, 2021Peschken, 2023Thomas}, or cosmological zoom-in simulations \citep{2019Buck, 2020HopkinsII, 2021HopkinsII, 2023MartinAlvarez, 2024RodriguezMontero}. Complementary to these ``full-physics'' approaches, a recent body of work studies the influence of CRs on the acceleration and thermodynamical evolution of cold clouds in the CGM \citep{2018Butsky, 2022Huang, 2023Tsung, 2024Armillotta}. Thematically aligned to CR-focused simulations, the ever-increasing resolving power of present-day simulations allows for exploring the magnetic field evolution and amplifying dynamo processes on galactic scales \citep{2009Wang, 2017Rieder, 2017Pakmor,2021Whittingham, 2022Pfrommer, 2022Wibking, 2023Whitworth, 2024Pakmor}. 

An observation derived from these simulation efforts is that CR transport significantly changes the strength of CR feedback. Guided by this finding, new hydrodynamical models have been developed \citep{2018Jiang, 2019Thomas}, which include more information about the momentum-space structure of the CRs to allow for a more detailed and realistic description of CR transport. The aforementioned simulations routinely employ a gray (or mono-energetic) approximation for the description of the entire CR population and solve for the CR flux density in addition to the energy density, i.e., they account for the isotropic and first-order anisotropy of the CR distribution function while still assuming quasi-linear theory of CR transport. This gray approach is justified by the observation that, under typical circumstances, most of the CR energy is carried by CR protons with GeV energies. 

However, this approach neglects the multi-faceted nature of CR transport and might produce inaccurate answers for CR-ionization in molecular clouds, which is mediated by MeV CR protons, or for the $\gamma$-ray emission originating from TeV CR protons. Moreover, the dynamical impact of the CRs is affected through energy-dependent CR transport speeds, in particular in the CGM \citep{2020Girichidis, 2020Hopkins,  2021Ogrodnik, 2022Girichidis, 2024Girichidis}. A tight connection between CRs and star formation is observed through the emission of radio-synchrotron radiation \citep{2003Bell} and $\gamma$-rays \citep{2012Ackermann} which correlate with the infrared-emission. Analysing simulations with respect to these emission mechanisms opens up an avenue to interpret and compare with observations \citep{2017PfrommerII,2021WerhahnI,2021WerhahnII,2021WerhahnIII,2023Werhahn,2023Boess,2024Ponnada}.

In the present work, we are outlining the quantitative differences between galactic winds that are driven with and without the additional support of CRs. To achieve this, we are comparing a pair of simulations of isolated Milky Way-type disk galaxies run with the \textsc{Arepo} moving-mesh code accounting for relevant microphysical cooling and heating physics together with stellar feedback through the \textsc{Crisp} (Cosmic Rays and InterStellar Physics) framework. Our simulations employ a super-Lagrangian refinement strategy that targets the inner CGM of the simulated galaxy and tremendously increases the numerical resolution in this region. This enables us to investigate the wind launching processes and the resulting thermodynamical phase structure of the wind in great detail. We shall show that the presence of CRs increases the mass loading of the simulated galactic wind and lowers the temperature of the outflow to an extent that observable column density of \ion{O}{VI} and \ion{C}{IV} are increased by more an order of magnitude. 

The outline of this paper is as follows. In Section~\ref{sec:setup}, we describe our simulation approach, the isolated galaxy setup and the physics accounted for in the \textsc{Crisp} model. In Section~\ref{sec:outflow}, we give a visual overview of the simulated galaxy and the emerging galactic winds. The realized mass and energy loading factors inside the galactic outflows are analysed in Section~\ref{sec:loading_factors}. In Section~\ref{sec:wind_driving}, we investigate how the galactic winds are driven and at which galactic height they are accelerated. To support the idea of a CR-mediated galactic outflow, we compare vertical pressure profiles of both galaxies in Section~\ref{sec:pressure_profiles}. The presence of a global CR pressure gradient alters the thermodynamical phase structure of the galactic winds which we analyse in Section~\ref{sec:phase_structure}. Lastly, in Section~\ref{sec:metal_absorption}, we show that the altered phase structure has direct consequences for metal line absorption properties.

\section{Simulation setup}
\label{sec:setup}

The simulations presented in this paper are conducted with the moving-mesh code \textsc{Arepo} \citep{2010Springel,2016PakmorII} and employ the magneto-hydrodynamics (MHD) module \citep{2011Pakmor, 2013Pakmor}, and the CR module \citep{2017Pfrommer} with its extension for 2-moment CR-magneto-hydrodynamics \citep[CRMHD,][]{2019Thomas, 2022Thomas, 2021Thomas}. We model the evolution of the galaxy using the \textsc{Crisp} framework. \textsc{Crisp} aims to account for relevant microphysical processes that impact the ISM, follows the transport of CRs in the ISM and CGM, and addresses how CR and stellar feedback shape the gas dynamics of a galaxy and hence its star formation capabilities. In this section, we present a short summary of the \textsc{Crisp} framework while a complete presentation and discussion are given in our companion paper (Thomas et al., in prep.). 

We compare two simulations sharing the same initial conditions, both including the galactic and ISM physics described by the \textsc{Crisp} framework. The sole difference between the two simulations is that one models the evolution of the galaxy without CR physics. This simulation run is named MHD in the following. The other simulation run includes the full set of CR physics and dynamics described below and is called CRMHD in the following.

\subsection{ISM model}

\textbf{Star formation} is modelled in the standard way by statistically converting gas cells into star particles once the gas is assumed to be dense and collapsed. We assume that this process can be described using the standard Schmidt-type approach with a star formation efficiency per free-fall time given by
\begin{align}
    \epsilon_\mathrm{ff} = 
    \begin{cases}
    100\% & \mathrm{for} \quad \rho > 100 \, m_\mathrm{p}~\mathrm{cm}^{-3} X^{-1}, \\
    0\% & \mathrm{overwise}, \\
    \end{cases}
\end{align}
where $X=0.76$ is the mass fraction of hydrogen. Hence, we consider gas above 100 hydrogen particles per cm$^{3}$ as star-forming. We choose this threshold because gas at higher densities is most likely situated in molecular clouds, whose internal dynamics and physical processes we cannot reliably resolve at our resolution of $1000~\mathrm{M}_\odot$. High star-formation thresholds are generally needed to bring the clustering of newly formed star particles in agreement with the observed strong clustering of newly born stars \citep{2019Buck,2022Keller}.This high efficiency guarantees that gas above this threshold is quickly converted into stars -- on the free-fall time, which is $t_\mathrm{ff} = (3 \pi / 32 G \rho)^{1/2} = 4.5$~Myr at the threshold. Therefore, the threshold is an effective upper limit of the density reached in our simulation.

After creating a star particle, it starts participating in stellar feedback processes. At our target resolution, we are not resolving individual stars but rather single stellar populations whose feedback yields are averaged over a distribution of stellar masses. First, stellar winds and later, core-collapse type II SN explosions of massive stars expel mass, metals, and energy into their surrounding media. Their respective rates are based on \textsc{Starburst99} calculations \citep{1999Leitherer}. The energy released by SNe is injected into the ISM either using thermal energy or momentum carried by the SN shell. Which of the two forms we use depends on whether the Sedov-Taylor phase of an SN can be resolved. SN and stellar winds form shocks at which CR can be accelerated and later escape into the galaxy. We model this by injecting 5\% of the mechanical energy delivered by the stellar wind and the SN explosion in the form of CR energy. We sample individual SNe following the to the \textsc{Starburst99} rate and inject the released $1.06\times10^{51}~\mathrm{erg}$ energy for each SN event into an injection region around the star particle. The size of the injection region is chosen so that there is a layer of two gas cells in each direction around the star particle. Because we adopt a maximum cell size of $\Delta x=100$~pc within the galaxy, this implies a ceiling of 200~pc for the injection region. If required, we inject momentum in a conserving and isotropic manner to avoid potential errors introduced by mass bias \citep[see][for a discussion on the necessity of such a weighting scheme]{2018Smith}. A detailed description of the injection routine will be presented in Thomas et al. (in prep).

We follow 12 different species (H$_2$, \ion{H}{i}, \ion{H}{ii}, all ionization stages of He, and the first two ionization stages of C, O, and Si) inside the \textbf{chemistry module}. Their evolution is determined by radiative and collisional ionization, radiative and dielectric recombination, charge exchange reactions, dust-mediated processes, and CR ionization. Our modelling of the chemical state allows us to follow the thermodynamical state of the ISM accurately.

We include \textbf{cooling} by low-temperature fine-structure metal lines from the first two ionization stages of C, O, and Si through a direct calculation based on collision rates \citep{2007Abrahamsson,2014Grassi}, cooling by high-temperature metal lines using interpolating tables pre-compiled with the \textsc{Chianti} code \citep[which include iron and other metal lines,][]{1997Dere}, cooling provided by rotation-vibrational lines of H$_2$ \citep{2021Moseley}, Ly$\alpha$ cooling by H, and bremsstrahlung cooling at high temperatures \citep{1992Cen}. \textbf{Heating} inside the ISM is dominated by photoelectric heating through the absorption of FUV photons on dust grains \citep{1994Bakes} and residual energy of CR ionization and cooling processes \citep{2017Pfrommer} aside from the energy injected by SNe and stellar winds. 

The energy density of the \textbf{FUV} radiation field is calculated based on a reserve ray-tracing approach that is combined with a tree infrastructure to merge distant FUV sources \citep{2019Grond,2021Wuensch}. This allows us to reduce the computational cost but also to account for the absorption of FUV radiation by dust grains inside the ISM. To this end, we assume that the dust abundance scales linearly with that of metals. Radiation is emitted by young stars (age $<3\,$Myr) for which we calculate the luminosity based on the \textsc{Bpass} model \citep[Binary Population and Spectral Synthesis code by][]{2018Stanway}. 

Additional ionization of dilute gas in our simulations is provided by a constant \textbf{meta-galactic UV background}. We take the spectral radiation energy density from \citet{2019Puchwein} at redshift $z=0$ and calculate photoionization and heating rates based upon the cross-sections of \citet{1996Verner}. We assume that dense gas will be self-shielded from this radiation field and use the prescription of \citet{2013Rahmati} to attenuate the meta-galactic UV background.

\subsection{Cosmic-ray physics}
\textbf{CRs} are transported along magnetic field lines in an anisotropic manner using the CRMHD theory of \citet{2019Thomas}. We are following the entire population of CR protons in a gray approach and average their interaction rates using the spectrum of \citet{2017Pfrommer} weighted by the particle energy. Because CR protons with energies around 1 GeV dominate the energy budget of the overall CR population, our description is biased toward their evolution. The CR transport velocity is set by interactions of CRs with small-scale Alfv\'en waves, which exchanges energy through the gyroresonant instability \citep{1969Kulsrud,2023Shalaby}. These Alfv\'en waves can be damped through the non-linear Landau \citep{1991Miller} effect or through ion-neutral damping \citep{1969Kulsrud}. In addition to interactions with Alfv\'en waves, CRs lose energy to their surrounding plasma (partly in the form of heat) through hadronic and Coulomb interactions \citep{2017Pfrommer}.  

\textbf{Ion-neutral damping} operates because neutral and ionized particles of the ISM collide and exchange energy with one another \citep{1969Kulsrud}. As the magnetic field of the small-scale Alfv\'en waves is frozen into the motion of the ionized particles, they also lose some energy due to these collisions. The corresponding damping rate of the small-scale Alfv\'en waves is related to the momentum-transfer cross-section of the ion-neutral collisions \citep{2019Thomas}. We use the momentum-transfer cross-sections of \citet{2008Pinto} and only account for the leading-order interactions that contain at least an ionized or a neutral hydrogen atom or molecule as a collision partner. The fractions of neutral and ionized species are calculated from the output of the thermochemistry module.

\subsection{Initial conditions }

The presented simulations start from an \textbf{idealized setup}. We initialize a doubly exponential gaseous disk with a radial scale length of $5$~kpc, a scale height of $0.5$~kpc, and a total mass of $0.8\times10^{10}~\mathrm{M}_\odot$. The stellar disk has the same scale length and height but a total stellar mass of $3.2\times10^{10}~\mathrm{M}_\odot$. Star particles are initialized with a mass of $M_\mathrm{\star} = 1000~\mathrm{M}_\odot$ and are positioned randomly following the exponential disk profile. Their velocities are set by sampling a Maxwellian distribution whose velocity dispersions are calculated based on the Jeans equation following the approach of \citet{2005SpringelII}. Both stellar and gaseous disks are situated inside a static Hernquist halo with mass $M_{200} = 10^{12}~\mathrm{M}_\odot$, concentration $C=7$ and baryon mass fraction $f_\mathrm{b} = 0.155$ \citep{1990Hernquist}. The disk mass fraction is correspondingly $1/25$, of which $20\%$ are contained in the gaseous and $80\%$ are contained in the stellar component. 

The initial \textbf{magnetic field} consists of two components: one toroidal magnetic field which is dominant inside the galactic disk and a constant vertical magnetic field. For the toroidal magnetic field, we use a magnetic field strength of $B_\mathrm{tor} = 10^{-1} ~\mu\mathrm{G} \times \sqrt{\rho / \rho_\mathrm{max}}$, where $\rho$ is the gas density whose maximum is $\rho_\mathrm{max}$ at the centre of the galaxy. This scaling keeps the Alfv\'en speed constant and low throughout the initial galactic disk. The vertical component has a magnetic field strength of $10^{-3}~\mu$G. Both components are weak and memory about both components is quickly erased as the field strength is amplified by a small-scale dynamo once feedback and galactic shear affect the dynamics in and around the galaxy  \citep{2022Pfrommer}. 

\begin{figure*}
   \centering
   \resizebox{\hsize}{!} { \includegraphics[width=\textwidth]{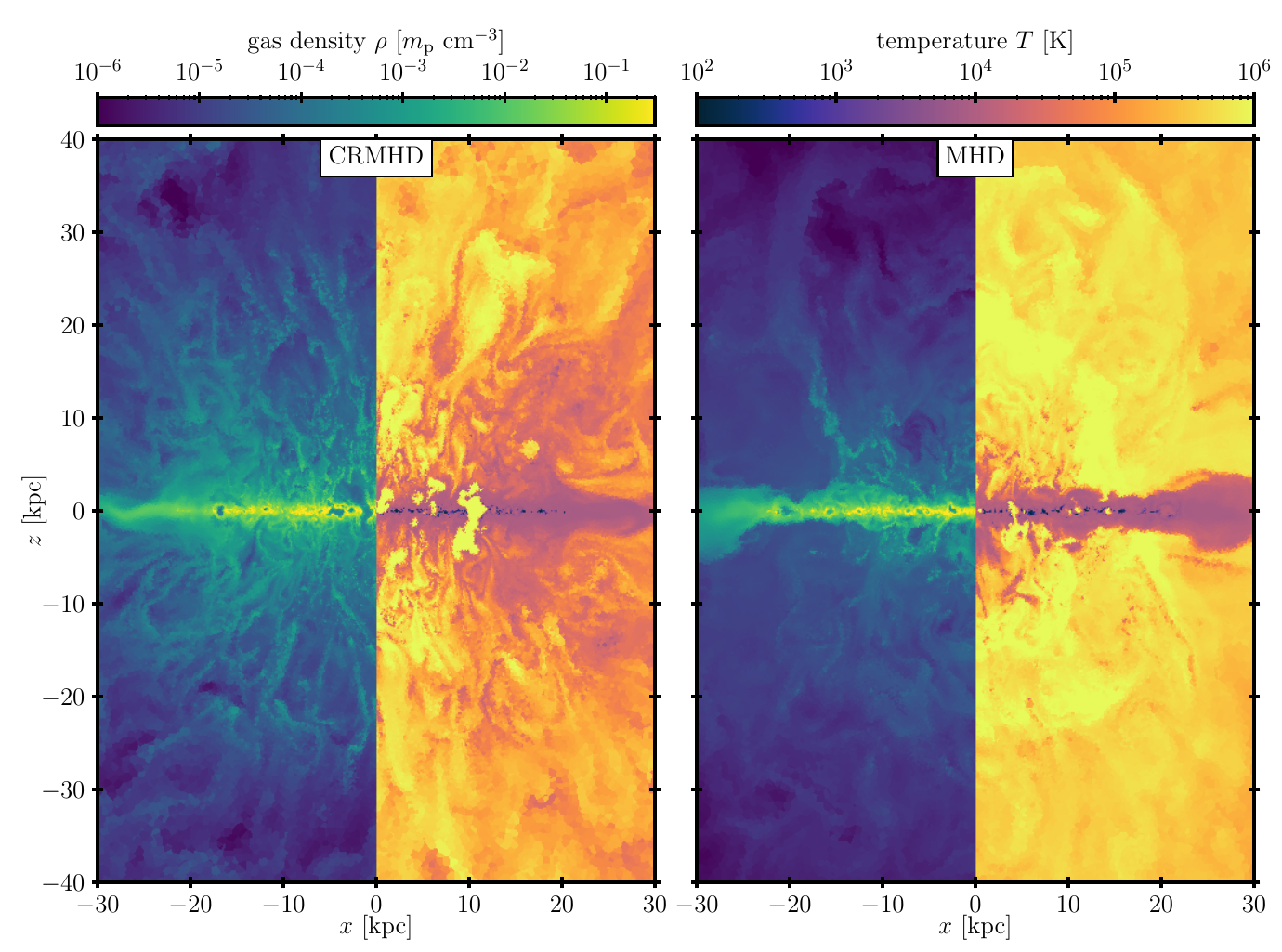} }
   \caption{Vertical slices through the galactic disks of the CRMHD (left) and the MHD (right) simulations at $t=1$~Gyr. Both panels show the hydrogen number density on their left and the gas temperature on the right half, respectively. Accounting for CR feedback results in a visually distinct inner CGM around the galaxy. The simulation which includes CRMHD physics has a denser and colder CGM and appears to be more turbulent in comparison to the MHD simulation.}
   \label{fig:gallery}
\end{figure*}

The \textbf{target resolution} in our simulation is $M_\mathrm{target} = 1000~\mathrm{M}_\odot$ and gas cells that are at least twice as massive are refined to enforce this criterion again. Gas cells with masses lower than half the target resolution are derefined by merging them with their neighbours \citep{2010Springel}. On top of this, we employ super-Lagrangian refinement inside the galactic wind and the disk which tremendously increases the numerical resolution in this region at a moderate increase of required computational resources \citep{2019vandeVoort,2023Thomas,2024Rey}. This is realized in our simulation by defining shells with radius $r_\mathrm{shell}$ and requiring a maximal cell size $\Delta x_\mathrm{max}$ for the computational elements inside these shells. Gas cells that reside inside these shells and have volumes $V > 4\pi / 3  \Delta x_\mathrm{max}^3$ will be refined to enforce this resolution target where we assume that Voronoi cells are close to spherical to define an equivalent cell size. We use 4 nested shells around the galactic centre:
\begin{itemize}
    \item $r_\mathrm{shell,1} = 15$~kpc with $\Delta x_\mathrm{max}= 100$~pc,
    \item $r_\mathrm{shell,2} = 30$~kpc with $\Delta x_\mathrm{max}= 200$~pc,
    \item $r_\mathrm{shell,3} = 60$~kpc with $\Delta x_\mathrm{max}= 400$~pc,
    \item $r_\mathrm{shell,4} = 120$~kpc with $\Delta x_\mathrm{max}= 1000$~pc
\end{itemize}
to properly resolve the dynamics inside the galactic wind and the disk, and to increase the number of cells in this region which will become useful for our later statistical analysis.

\begin{figure*}
   \centering
   \resizebox{\hsize}{!} { \includegraphics[width=\textwidth]{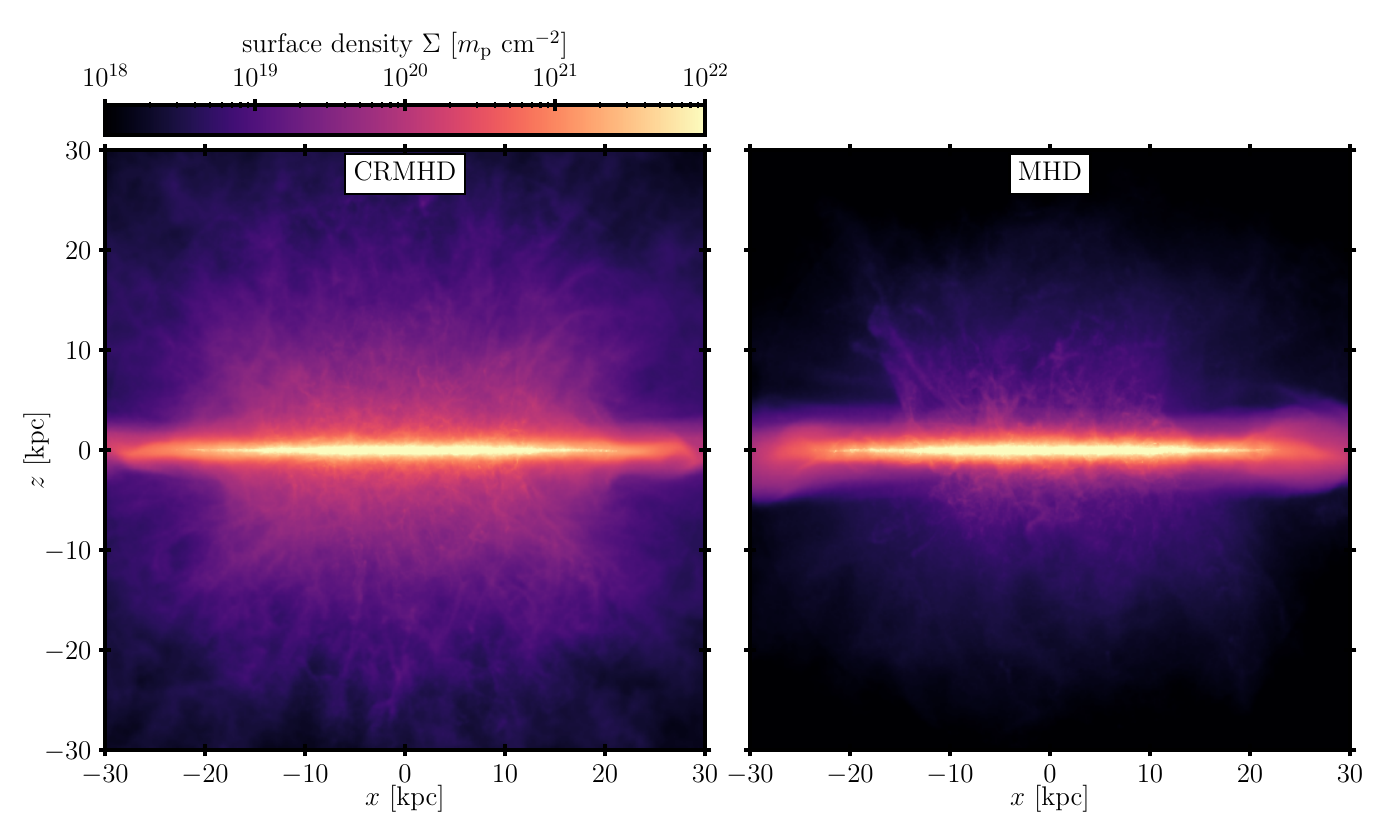} }
   \caption{Vertical surface densities projected through the galactic disks and inner CGM of both the CRMHD (left) and the MHD (right) simulations at $t=1$~Gyr. Stellar feedback and CRs lift significantly more gas mass above the CRMHD disk than the much weaker galactic wind originating from the MHD galaxy, which lacks the additional support by CRs. }
   \label{fig:gallery_3}
\end{figure*}

\begin{figure*}
   \centering
   \resizebox{\hsize}{!} { \includegraphics[width=\textwidth]{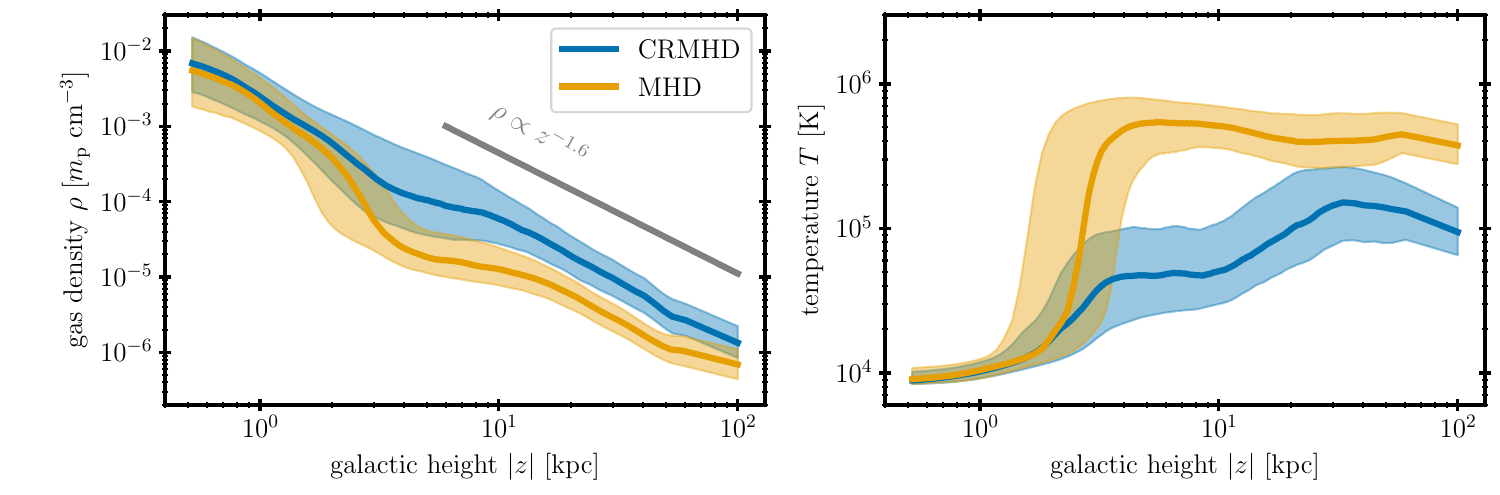} }
   \caption{Vertical profiles of the hydrogen number density and gas temperature measured above the galactic disk inside a cylinder with radius $R < 30$~kpc. We show the median as a thick line and the shaded area spans from the 20$^\mathrm{th}$ to 80$^\mathrm{th}$ percentile of the volume-weighted quantity. Both, the CRMHD and MHD galaxies show similar profiles near the galactic disk. Above 2-3 kpc, their profiles start to differ and the MHD simulation shows a clear phase transition with hotter and more dilute gas at larger galactic heights.}
   \label{fig:density_temperaure_profile}
\end{figure*}

\begin{figure*}
   \centering
   \resizebox{\hsize}{!} { \includegraphics[width=\textwidth]{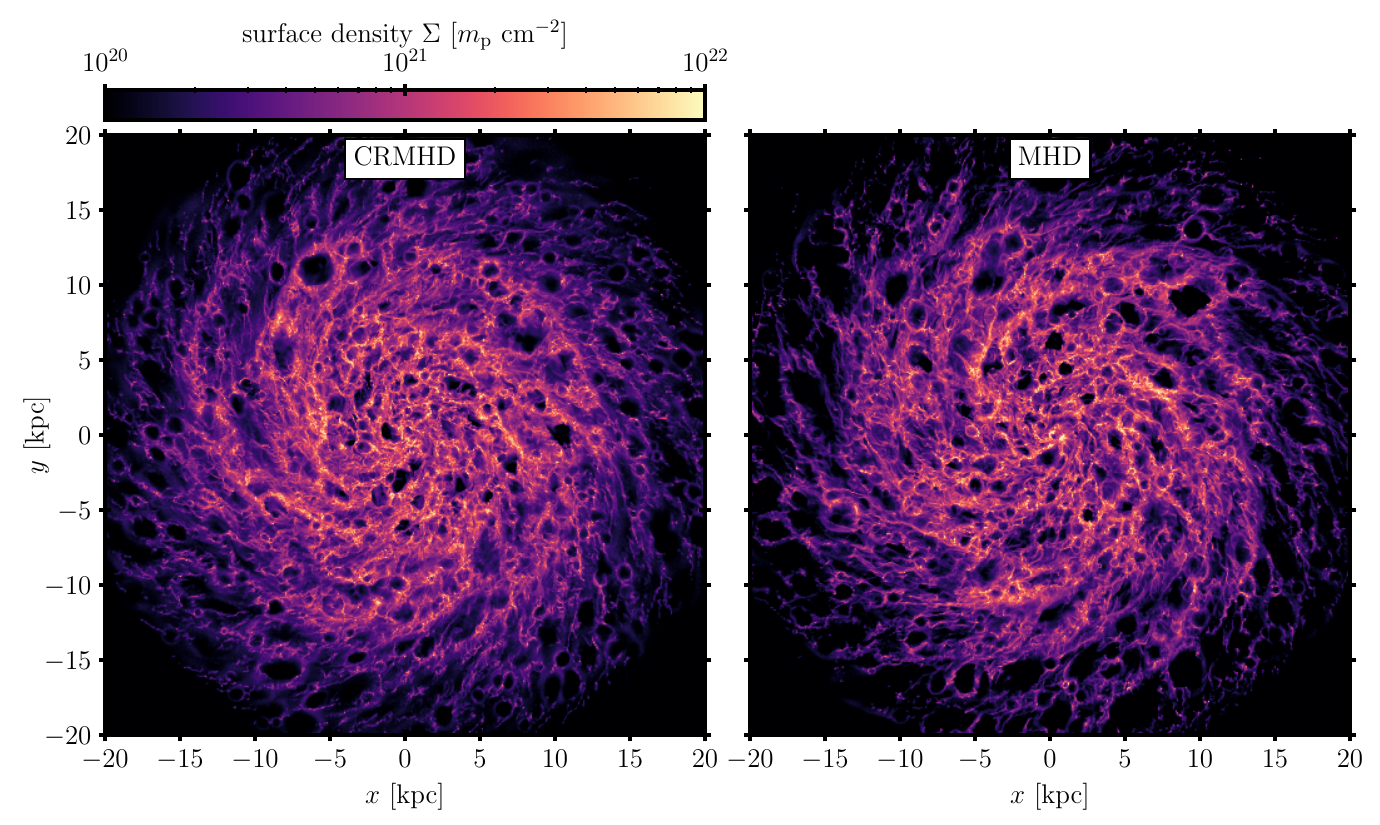} }
   \caption{Hydrogen surface densities of the CRMHD (left) and the MHD (right) simulations at $t=1$~Gyr. Both simulations show morphologically similar structures inside their ISM, which consists of various molecular clouds connected by filamentary dense ISM gas in addition to feedback-driven (super-)bubbles rupturing the galactic disk. There is no visually apparent indication of an increased star formation and feedback activity in either of the simulations, which could readily explain the differences in the galactic winds seen in Fig.~\ref{fig:gallery}. }
   \label{fig:gallery_2}
\end{figure*}

\section{Galactic disk \& outflow}
\label{sec:outflow}

\begin{figure}
   \resizebox{\hsize}{!} { \includegraphics[width=\textwidth]{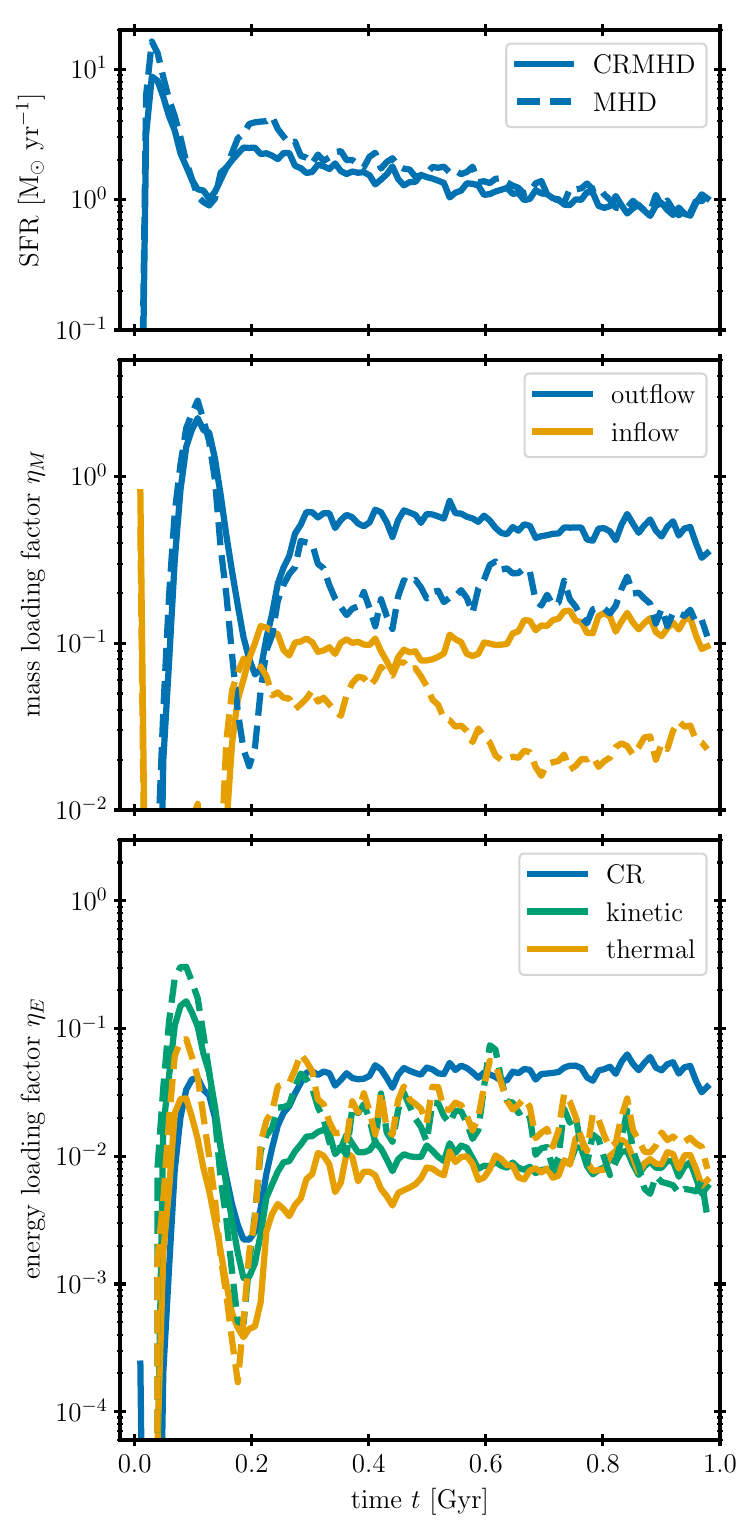} }
   \caption{Evolution of the SFR, the mass loading factors of outflow and inflow of gas, and the energy loading factors of the kinetic and thermal energy of the gas as well as the energy loading factor of the CRs. We calculate the mass loading factor based on Eq.~\eqref{eq:mass_loading} and the energy loading based on Eq.~\eqref{eq:energy_loading}. While both simulations share similar SFRs, the outflow rate and the energy loading is increased in the simulation accounting for CRs.}
   \label{fig:loading_factors}
\end{figure}

\begin{figure*}[h!]
\begin{minipage}[t]{.49\textwidth}\vspace{0pt}
   \includegraphics[width=\textwidth]{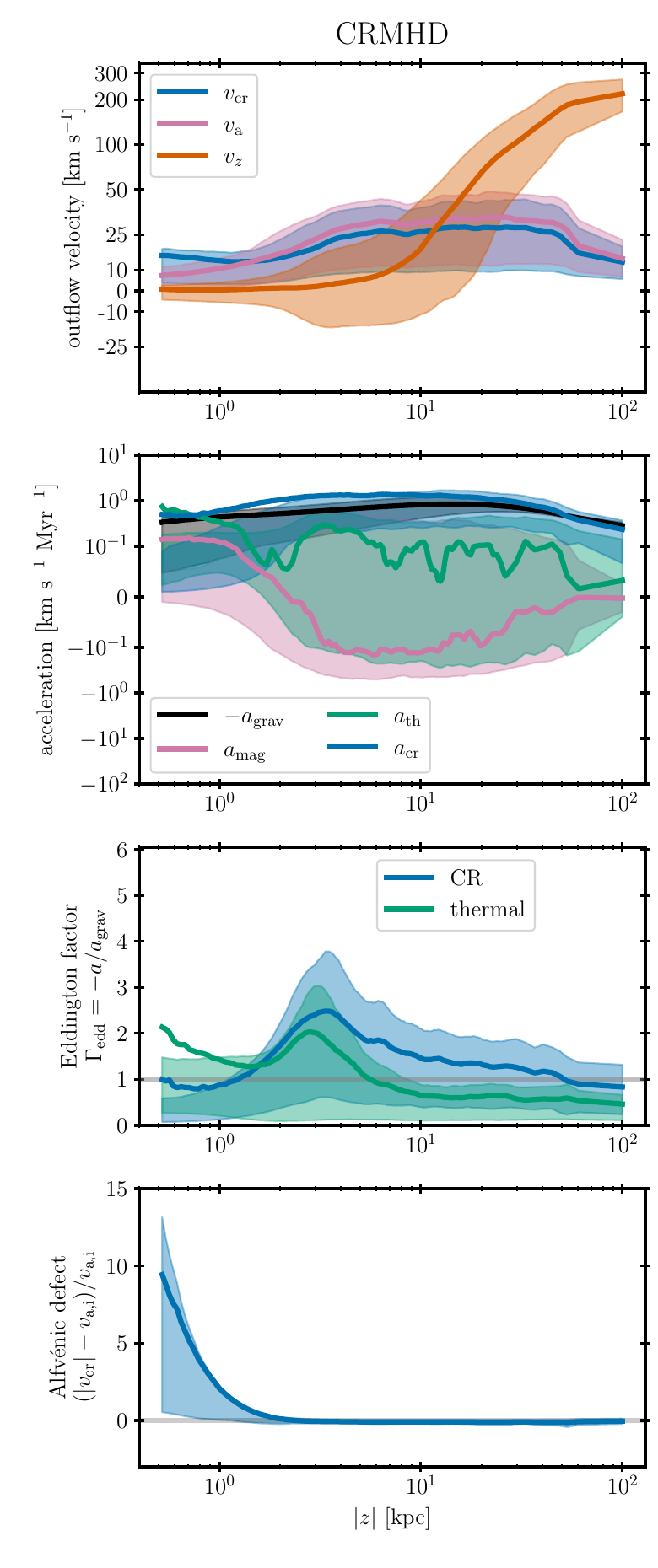}
\end{minipage}
\hfill
\begin{minipage}[t]{.49\textwidth}\vspace{0pt}
   \includegraphics[width=\textwidth]{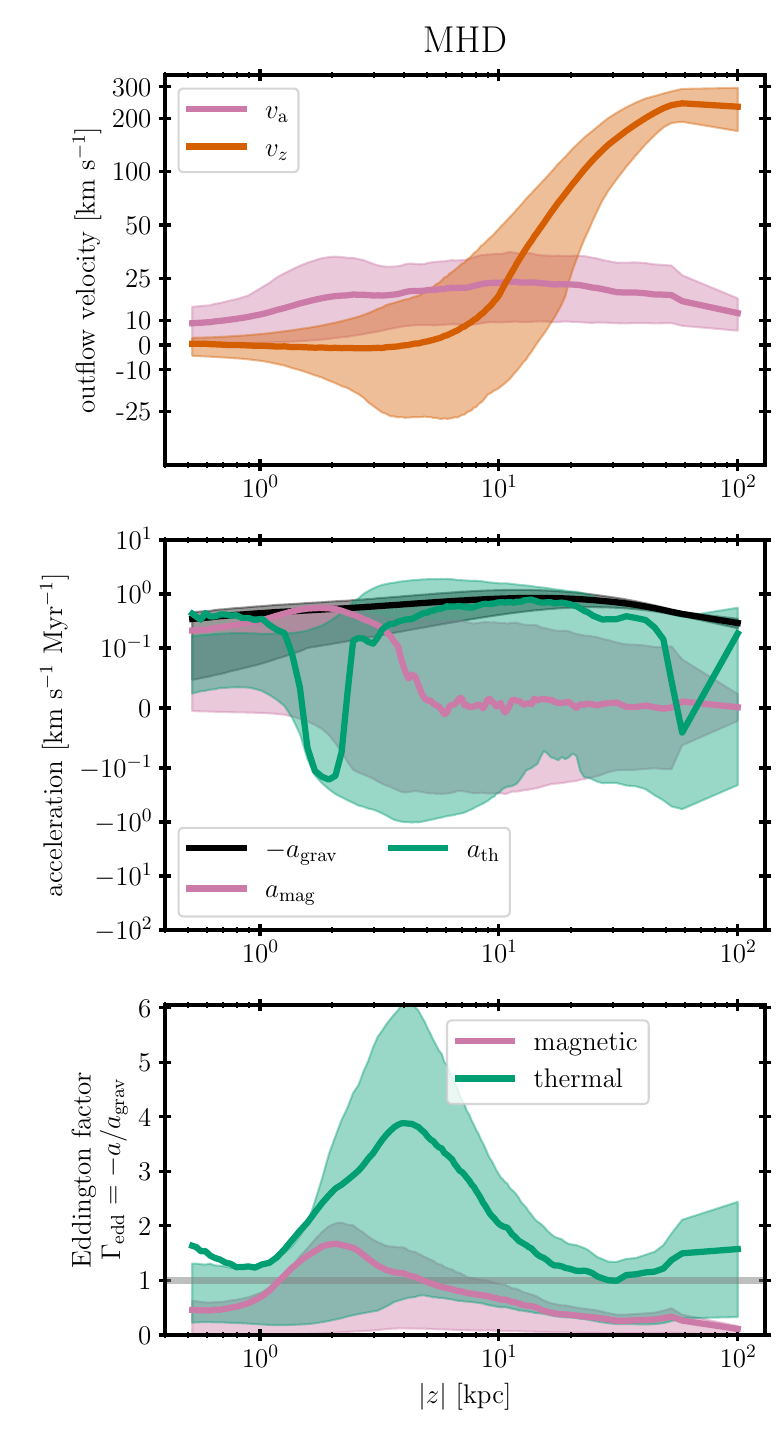}
   \caption{Vertical profiles of various outflow properties inside a galactocentric radius of $R<30$~kpc. The thick line in each panel shows the mean value at each galactic height while the shaded area spans from the 20$^\mathrm{th}$ and 80$^\mathrm{th}$ percentile of the volume-weighted quantity. The top panels show the vertical outflow velocity of the gas, the central panels show the actual value of the relevant accelerations $a$ as well as their Eddington factors $\Gamma_\mathrm{edd} = - a / a_\mathrm{grav}$ where $a_\mathrm{grav}$ is the gravitational acceleration, and the bottom panel shows the Alfv\'enic defect, which we define in and discuss after Eq.~\eqref{eq:alfenic_defect}. Both galactic winds are first accelerated at galactic heights of 2-3 kpc where the forces of the CRs in the CRMHD simulation or thermal and magnetic forces in the MHD simulation overcome gravity.}
   \label{fig:force_balance}
\end{minipage}
\end{figure*}

We use this section to briefly give an overview of the simulated galaxies, their ISM, and their galactic winds. 

In Fig.~\ref{fig:gallery}, we show vertical slices through the inner CGM and the galactic disk of both galaxies. We display the hydrogen number density and the temperature for both galaxies at the end of our simulation at $t=1$ Gyr. At this time, the general morphological structures of the wind reached a steady state and are not influenced by the initial starburst anymore. The densities and the temperature maps show different morphological features and values at comparable heights above the galactic disk. The inner CGM in the CRMHD simulation is generally colder and denser than the inner CGM of the MHD simulation. 

In the CRMHD simulation, we observe many interconnected cloudlets with elongated shapes that have an apex pointing toward the galactic plane. Gas inside the cloudlets is colder and denser than the dilute gas filling the space in between them. Clustered feedback originating from multiple SNe with a possible common origin inside the same GMC drives high-temperature (super-)bubbles inside the galactic disk which are able to break out of the disk and rise into the CGM. Some of the resulting high-temperature post-breakout plumes are visible near the galactic disk and a possible origin for the high-temperature gas between the cloudlets. The origin of the cloudlets is not directly apparent. A potential formation scenario is that the superbubbles sweep up dense ISM gas and carry it into the inner CGM or thermal instability in the CGM.

In the MHD simulation, the inner CGM appears to consist of multiple bipolar plume structures that rise from the galactic disk and are filled with hot gas with temperatures $\sim10^{6}$~K. Close to the galactic disk, individual dense and cold filaments are visible but do not extend to larger galactic heights. Moreover, they are not as volume filling as the network of cloudlets in the CRMHD simulation. A possible explanation for the morphological appearance of the dense gas distribution is its participation in a fountain flow so that the dense gas will eventually fall back onto the galactic disk. 

Although we are employing a super-Lagrangian refinement scheme to increase the numerical resolution inside the region shown, we are not resolving the cooling and mixing dynamics of the hot and cold media in the inner CGM. To properly resolve the cooling length, a sub-pc resolution is required \citep{2018McCourt,2019Sparre,2020Fielding} which is prohibitively expensive for our simulation which follows the evolution of a global galaxy. Nevertheless, the cloudlet morphology observable in CRMHD is reminiscent of the ISM simulation results \citep{2016Girichidis,2023Simpson}, which use a tallbox simulation setup and are consequently able to better resolve the cloudlets and their mixing with the hot phase of the outflow. 

In Fig.~\ref{fig:gallery_3}, we show the surface densities of gas in the inner CGM around the galaxies. In both simulations, this medium is interspersed by filamentary features that overlap in projection and create the impression of smooth inner CGM. The major difference between the CRMHD and MHD simulations is the increased amount of material that is entrained to the CGM around the inner galaxy in the former simulation. Furthermore, the density of filamentary gas tails appears to be higher around the CRMHD galaxy while these features are more concentrated at lower galactic heights for the MHD galaxy.

To quantify the different distributions of density and temperature of the outflow, we show their vertical profiles for both simulations in Fig.~\ref{fig:density_temperaure_profile}. We display the median as well as the 20$^\mathrm{th}$ and 80$^\mathrm{th}$ percentile of both volume-weighted quantities. At low galactic heights of $|z| < 2$~kpc, both galaxies show similar density and temperature profiles corresponding to the band of $10^4$~K gas that builds up the outer portion of the galactic disks and is visible in Fig.~\ref{fig:gallery}. At $|z| \sim 2$--$3$~kpc both galaxies show a transition that is most pronounced in the MHD simulation. The gas density drops by around an order of magnitude and attains temperatures of $5\times10^5$~K in the MHD galaxy. The gas in the CRMHD simulation only increases to maximum temperatures of around $5\times10^4$~K while the gas density only slightly flattens. In the simplified picture of a galactic fountain flow, dense and cold material expelled from the galaxy will fall back to the galaxy after reaching a maximum galactic height if it is not sufficiently accelerated to leave the galaxy. Above this maximum galactic height, such material thus cannot be found. This mere absence of dense and cold gas at larger galactic heights is thus a possible reason for this apparent phase transition. We will later discuss that both galactic winds are launched at this transition point.

At larger galactic heights, gas inside the outflow decompresses and shows a power law-like decline with galactic height following $\rho \sim z^{-1.6}$. The temperatures in both simulations remain rather stable and high -- only the CRMHD simulation shows an increase in temperature at $z\sim30$~kpc to $T \sim 10^{5}$~K. Maybe due to a coincidence but interestingly enough the median gas temperatures of both simulations are on the opposing sides of the peak of the metal-line cooling curve at around $2-3\times10^5$~K which is caused by the doublet lines of \ion{O}{vi}.

In Fig.~\ref{fig:gallery_2}, we show face-on projections of the gas density, i.e., the surface density for both simulations. Both galactic disks show a highly structured gas distribution that is composed of GMCs and regular molecular clouds that are interconnected by a filamentary almost spongiform network of dense gas. In between these overdensities, there is a variety of (super-)bubbles of different shapes and sizes that have been driven into the ISM due to recent stellar feedback events. Both surface density maps look similar with no clear sign that either of the galaxies has been experiencing noticeably more star formation and stellar feedback events. This would be visible as an over-abundance or possibly larger feedback bubbles. Furthermore, star formation and feedback are distributed uniformly throughout the galactic disks with some indication for locally enhanced rates due to the presence of spiral arms. We analyse the star formation rate (SFR) quantitatively in the next section.

\section{Energy and mass loading}
\label{sec:loading_factors}

In this section, we compare our two simulations by investigating their differences in terms of the global SFR as well as mass-, and energy-loading factors of the galactic wind. 

In Fig.~\ref{fig:loading_factors}, we display the evolution of these quantities. We analyse the loading factors in a cylindrical region with radius $R<20~$kpc at a galactic height of $10~\mathrm{kpc} < |z| < 11~\mathrm{kpc}$ which has a height of $\Delta z = 1~$kpc. The mass loading factor $\eta_{M}$ is defined as the mass flux through this region and is scaled to the instantaneous SFR via:
\begin{equation}
    \eta_{M} = \frac{1}{\mathrm{SFR}} \sum_{\mathrm{cells}~i} \frac{M_i \varv_{\mathrm{in/out},\,i}}{\Delta z}, \label{eq:mass_loading}
\end{equation}
where $M_i$ is the gas mass of the cell, $\varv_{\mathrm{in/out,}\,i}$ is the in- and outflow velocity with respect to the vertical $z$-axis:
\begin{equation}
    \varv_{\mathrm{in/out,}\,i} = (\vec{\varv}_i \bcdot \Vec{e}_z) \mathrm{sign}(z),
\end{equation}
which we calculate based on the gas velocity $\vec{\varv}_i$.

To calculate the outflow or the inflow mass loading factors, the sum extends over computational cells with positive or negative $\varv_{\mathrm{in/out,}\,i}$, respectively. The energy loading factors are scaled to the SN energy that is expected to be released by the current SFR (which is assumed to be $10^{51}~$erg per 100 $\mathrm{M}_\odot$ of newly formed stars) and is thus calculated by:
\begin{equation}
    \eta_{E} = \left( 10^{51} \mathrm{erg}~ \frac{\mathrm{SFR}}{100~\mathrm{M}_\odot} \right)^{-1} \times \sum_{\mathrm{cells}~i} \frac{E_i \varv_{\mathrm{out},\,i}}{\Delta z}, \label{eq:energy_loading}
\end{equation}
where $E_i \varv_{\mathrm{out},\,i}$ is the energy flux of a given computational cell, we use
\begin{align}
    E_i &= \frac{M_i}{2} \varv_i^2 \quad \text{for the kinetic energy flux,} \\
    E_i &= V_i (\varepsilon_{\mathrm{th,}\,i} + P_{\mathrm{th,}\,i})  \quad \text{for the thermal energy flux,}
\end{align}
and $\varepsilon_{\mathrm{th,}\,i}$ is the thermal energy density, $P_{\mathrm{th,}\,i}$ is the thermal pressure, and $V_i$ is the volume of the computational cell. We assume an adiabatic index of $5/3$ for the thermal pressure and $4/3$ for the CR pressure. The SFR is averaged over 10 Myr of evolution. CRs are not only advected with the gaseous flow but are also transported independently along magnetic field lines. We account for this additional flux and use an extended definition for the CR energy loading factor:
\begin{equation}
    \eta_{E,\,\mathrm{cr}} = \left( 10^{51} \mathrm{erg}~ \frac{\mathrm{SFR}}{100~\mathrm{M}_\odot} \right)^{-1} \times \sum_{\mathrm{cells}~i} \frac{E_{\mathrm{cr,}\,i}\varv_{\mathrm{out},\,i} + F_{\mathrm{cr,}\,i}}{\Delta z},
\end{equation}
where the additional flux is
\begin{equation}
    F_{\mathrm{cr,}\,i} = V_i f_{\mathrm{cr,}\,i} (\vec{b} \bcdot \vec{e}_z)  \mathrm{sign}(z),
\end{equation}
where $f_{\mathrm{cr,}\,i}$ is the CR energy flux density and $\vec{b} = \vec{B} / B$ is the direction of the magnetic field while we use
\begin{align}
    E_{\mathrm{cr,}\,i} &= V_i (\varepsilon_{\mathrm{cr,}\,i} + P_{\mathrm{cr,}\,i}),
\end{align}
for the advective CR flux where $\varepsilon_{\mathrm{cr,}\,i}$ is the CR energy density, $P_{\mathrm{cr,}\,i}$ is the CR pressure. In the top-left panel of Fig.~\ref{fig:force_balance}, we show both the vertical gas velocity (i.e., the velocity at which CRs are advected) and the CR transport velocity along magnetic field lines. At low galactic heights, the transport along magnetic field lines is faster than the advective transport while the opposite holds true at larger galactic heights above the launching point of the galactic wind. At the height of $\sim 10$~kpc both are equally important. Consequently, both transport mechanisms need to be taken into account for an accurate analysis of the CR energy loading factor.

Both simulations go through an initial starburst at $t\sim 100~$Myr where the SFR peaks at $\sim10~\mathrm{M}_\odot~\mathrm{yr}^{-1}$. This is followed by an outflow episode with high mass and energy loading factors where the energy of the outflow is mostly carried by kinetic energy. The galactic winds are thus ballistically driven as expected for a starburst. Both galaxies share mostly the same evolution during this phase and only the levels of SFR and energy loading in the MHD simulation are slightly elevated in comparison to the CR simulation. The energy loading during this phase reaches values at around 15\% while peaking at 30\% for the MHD simulation. Similar values for energy loadings are expected for starbursts and have been seen in simulations which focus on this extreme regime of stellar feedback \citep{2020Schneider}. Following the starburst, the galaxies evolve at low SFR for $\sim 200~$Myr. During these times, no significant energy output is produced in the galaxy and gas is falling back onto the galaxy. After $t\sim300~$Myr, the galaxies enter a steady state with an SFR at around $1$--$2~\mathrm{M}_\odot~\mathrm{yr}^{-1}$ which has a declining trend throughout the rest of our simulation. Here, the MHD simulation shows a 15\% increase in SFR in comparison to the CRMHD simulation. This difference can be attributed to the presence of CRs in the ISM whose local pressure support can slow down collapse or provide a barrier offsetting compression from SNR. The difference between the SFRs gradually decrease towards the end of the simulation.

The galactic winds observed in Fig.~\ref{fig:gallery} start to emerge during this steady state phase. The outflow dominates the mass budget after $t\sim300~$Myr and reaches a mass loading factor of around $0.3$--$0.5$ for the CRMHD simulation and $0.1$--$0.2$ for the MHD simulation at late times. On average, the mass loading is increased by 167\% in the CHRD simulation.  The energy flux through the galactic wind is dominated by the CRs with energy loading factors of $0.3$--$0.5$ while most of the energy is carried by kinetic motions in the MHD simulation. The galactic wind in the MHD simulations carries overall more energy in the kinetic and thermal channels. The evolution of these energy loading factors fluctuates stronger in the MHD simulation. We attribute this to the overall more explosive nature of the galactic wind in the MHD simulation where strong superbubble breakouts can lead to a temporal increase in the energy loadings while CRs more smoothly drive the wind in the CRMHD simulation. In total, there is an overall 65\% increase in the energy that is transported by the galactic wind in CRMHD if we also account for the CRs. 

The CR energy loading factor settles at a value of around 4\%, which is quite comparable to the energy injection efficiency of 5\% at the supernovae itself. The CRs thus lose only a fraction of their total energy while escaping the ISM and moving into the CGM. Consequently, they retain their potential to provide feedback at larger galactic heights better than the direct thermal and kinetic energy of the SNe whose energy loading factors fluctuate at sub-percent levels. Hence, the majority of thermal SN energy and momentum is used for stirring turbulence either directly or by causing fountain flows, thereby self-regulating the ISM and not for efficiently driving outflows to large heights. Comparing both simulations, the larger mass and energy loadings of the galactic wind of the CRMHD simulation only have a minor influence on its instantaneous SFR rate. This is in line with our observation made in Fig.~\ref{fig:gallery_2} that both galaxies have a similar gas morphology and apparent impact of stellar feedback. The depletion of gas in the CRMHD galaxy is thus not regulated by the mass outflow but by star formation. We note that our idealized setup precludes robust predictions of the long-time behaviour of the SFR, which would necessarily have to include a cosmological setting.

\section{Wind driving}
\label{sec:wind_driving}

In this section, we investigate the origin of galactic winds by comparing the potential physical driver mechanisms and analyse the relevant forces acting on the gas. 

In the upper panel of Fig.~\ref{fig:force_balance}, we show the outflow velocity of the gas from the galactic disk (which is given by $\pm \varv_z$ depending on whether the gas cell resides at positive or negative $z$). We focus on a cylindrical region defined by $R < 30$~kpc and $|z| < 100$~kpc instead of taking the whole galactic wind into account. Our results are robust to changes of the analysis region. We again analyse the output of 10 simulation snapshots that are equally spaced between 900~Myr and 1~Gyr. The outflow pattern of the galactic wind is not quite radial and deviations from a pure radial or cylindrical geometry complicate and obfuscate the clear conclusion that we get from focussing on this region. Both simulations show similar velocity profiles. At low galactic heights, there is no acceleration of the galactic winds visible and the outflow velocity fluctuates at around 0. Here the galactic fountain flows shape the dynamics. This changes at $\sim 10$~kpc where the winds start to be launched and significantly accelerate at galactic heights between $10$ and $50$~kpc to eventually reach $\varv_z \sim200-300$~Km s$^{-1}$ at $\sim100$~kpc. At larger galactic heights, the wind transitions due to the presence of a wind termination shock.

The acceleration profiles in the middle panel of Fig.~\ref{fig:force_balance} help us to identify the drivers of the winds. We compare the main forces acting on the gas: the thermal pressure force $\vec{\nabla} P_\mathrm{th}$, the combined force from magnetic pressure and tension $\vec{\nabla} \bcdot (B^2 / 8 \pi \tens{1} - \vec{B} \vec{B} / 4 \pi) $, adopting Gaussian cgs units, as well as the force exerted by CRs $\vec{\nabla} P_\mathrm{cr}$.\footnote{We opt to use the full gradient force for the CR-mediated force instead of the more accurate one that differentiates between the CR-gradient force perpendicular to the magnetic field direction and the Alfv\'en-wave mediated force parallel to the magnetic field direction \citep{2019Thomas}. The full-gradient force is a better representation of the average CR force in the steady state which is of interest for the present analysis.}  We also plot the negative value of the gravitational force. For each of the forces, we only show its vertical component. To explain the gas acceleration away from the galactic disk, one of these forces or possibly a combination of those has to overcome the gravitational force. For the galactic wind in the CRMHD simulation, only the thermal and CR acceleration are strong enough to counter-balance gravity. For the MHD simulation, both the magnetic and the thermal forces are significant wind drives.  

Interpreting the actual values in the acceleration profiles is difficult due to the turbulent nature of the galactic winds. The average and region spanned by the 20$^\mathrm{th}$ to the 80$^\mathrm{th}$ volume-averaged percentiles fluctuate for all displayed quantities. This is not caused by low-number statistics because the average number of individual data points in our binning is around 300,000 but rather a representation of the volatile nature of this physical medium. Inside the galactic disk and in the inner CGM, averages of quantities might even show an excursion outside the region delineated by the 20$^\mathrm{th}$-80$^\mathrm{th}$ percentiles, which indicates that the statistics at these galactic heights are highly non-Gaussian.

To ease our analysis, we also compare the significantly accelerating components to the gravitational acceleration using the local Eddington factor $\Gamma_\mathrm{edd} = - a / a_\mathrm{grav}$ in Fig.~\ref{fig:force_balance}. If a given accelerating component can locally counteract gravity, then $\Gamma_\mathrm{edd} > 1$. For both simulations, the Eddington factor shows a peak at around galactic heights of 2-4 kpc where we previously saw that both galactic winds show apparent phase transitions; notably, an increase in temperature and a drop in density as seen in Fig.~\ref{fig:density_temperaure_profile}. Above this point, where the outflow velocities gradually increase, the Eddington factors decrease to the point where only the acceleration exerted by the CR pressure gradient in the CRMHD simulation and the thermal pressure force in the MHD simulation are able to drive gas away from the galactic disk. Consequently, they are responsible for driving the wind at larger galactic heights until the terminal speed of the galactic wind is reached. This finding in our CRMHD simulation is consistent with cosmological simulations, which find that the combined action of thermal and CR acceleration can accelerate hot gas around Milky-way analogues across cosmic time \citep[see Fig.~10 of][their analysis was done at $r \sim 0.2 R_\mathrm{vir}$]{2024RodriguezMontero}. 

\begin{figure*} 
   \centering
   \resizebox{\hsize}{!} { \includegraphics[width=\textwidth]{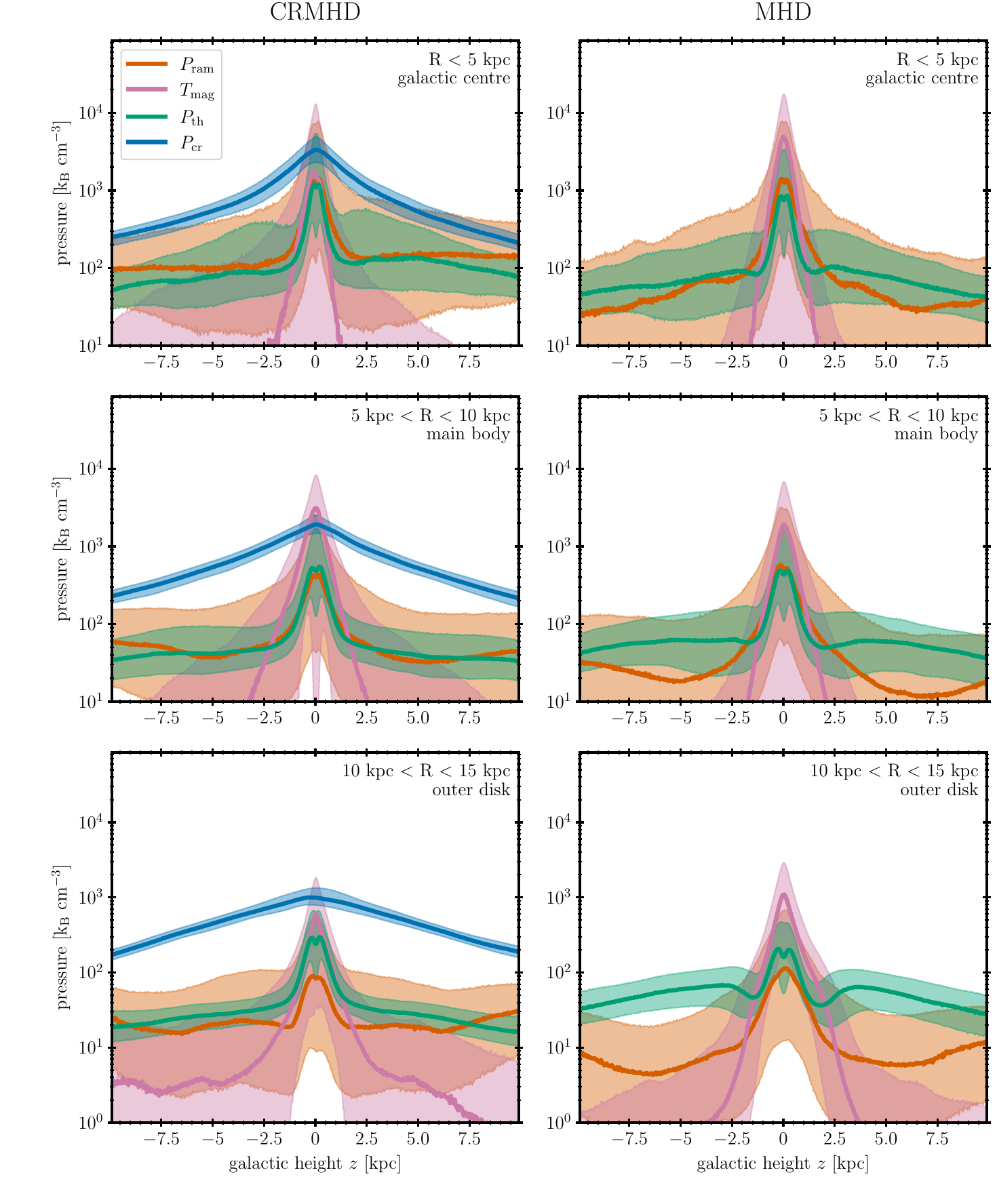} }
   \caption{Vertical pressure profiles through the galactic disk and the inner CGM of both the CRMHD (left) and MHD (right) simulations. We show the vertical ram pressure $P_\mathrm{ram} = \rho \varv_z^2$, vertical magnetic stress $T_\mathrm{mag} = (B_x^2 + B_y^2 - B_z^2) / 8 \pi$, thermal pressure $P_\mathrm{th}$, and CR pressure $P_\mathrm{cr}$. The respective thick lines show the volume-weighted median values at each galactic height while the shaded regions enclose the 20$^\mathrm{th}$ to 80$^\mathrm{th}$ volume-weighted percentiles. We differentiate between three regions: the inner galactic disk $R / \mathrm{kpc} < 5$ (top panels), the main body of the galactic disk $5 < R / \mathrm{kpc} < 10$, and the outer (still star-forming) galactic disk $10 < R / \mathrm{kpc} < 15$. Most non-CR pressure components show similar values if we compare the CRMHD and MHD simulations at similar galactic radii. Their values show a rapid decline from the galactic disk into the CGM. The CR pressure shows larger scale heights and is the dominant pressure component inside the inner CGM.  }
   \label{fig:pressure_profile}
\end{figure*}

To further demonstrate that CRs in fact drive the galactic wind in the CRMHD simulation, we now analyze the behavior of the
\begin{equation}
    \text{Alfv\'enic~defect} = \frac{|\varv_\mathrm{cr}| - \varv_\mathrm{a,i}}{\varv_\mathrm{a,i}},
    \label{eq:alfenic_defect}
\end{equation}
where $\varv_\mathrm{cr} = f_\mathrm{cr} / (\varepsilon_\mathrm{cr} + P_\mathrm{cr})$ is the effective transport speed of CRs, $\varv_\mathrm{a,i} = B / \sqrt{4 \pi \rho_\mathrm{i}}$ is the ion-Alfv\'en speed which is calculated based on the ion mass density $\rho_\mathrm{i}$. This quantity shows the mean velocity of the CR distribution with respect to the Alfv\'en waves. If this value is zero then CRs and Alfv\'en waves on average move at the same velocity and CRs behave according to the picture of streaming CR transport. Inside the ISM and below the point of the apparent phase transition, CRs are predominantly transported with super-alfv\'enic speeds. They efficiently couple into the movement of Alfv\'en waves at the point of the phase transition and continue to stream at larger galactic heights. There, they are efficiently coupled to the Alfv\'en waves, transfer momentum to accelerate the gas, and additionally heat the surrounding thermal gas. In our previous simulation of a CR-driven galactic wind using an ISM model with a pressurized equation of state, we also identified the point where CRs start to stream with the Alfv\'en waves to be critical for launching a galactic wind \citep{2023Thomas}.

\section{Vertical pressure profiles}
\label{sec:pressure_profiles}

To investigate the apparent phase transition and the wind-driving processes in more detail, we show vertical pressure profiles in Fig.~\ref{fig:pressure_profile}. We plot the median value in addition to the values spanning the 20$^\mathrm{th}$ to 80$^\mathrm{th}$ percentile of the volume-weighted pressure distribution. We conduct our analysis for three distinct annuli in our galaxies to distinguish regions with similar local SFR: the inner galactic centre is composed of gas with $R < 5$~kpc, the body of the galactic disk ranges from $5~\mathrm{kpc} < R < 10~\mathrm{kpc}$, and the outer disk is confined to $10~\mathrm{kpc} < R < 15~\mathrm{kpc}$. Their boundaries are chosen via visual identification in Fig.~\ref{fig:gallery_2} and each annulus spans a scalelength of the initial gas disk: the galactic centre shows little spiral structure, while the body and the outer disks separate the highly star-forming disk into an inner and an outer region. To increase the number of individual data points in our statistics, we include 10 equally-spaced snapshots between 0.9 and 1.0 Gyr for our analysis. We plot the pressure profiles and relevant pressures and stress quantities, namely the thermal pressure $P_\mathrm{th}$, the CR pressure $P_\mathrm{cr}$ and the vertical components of the magnetic stress tensor $\tens{T}_\mathrm{mag} = (B^2 / 8 \pi) \tens{1} - \vec{B} \vec{B} / 4 \pi$ which amounts to $T_{zz} \equiv T_\mathrm{mag} = (B_x^2 + B_y^2 - B_z^2) / 8\pi$ and the vertical component $P_\mathrm{ram} = \rho \varv_z^2$ of the ram pressure tensor $\tens{P}_\mathrm{ram} = \rho \vec{\varv} \vec{\varv}$. 

All profiles except for the CR pressure profile show different behaviours at low ($|z| < 2$~kpc) and high ($|z| > 2$~kpc) galactic heights. Both galaxies also show an apparent phase transition at the same galactic height. The pressures at low galactic heights and thus within the galactic disk show a small scale height and are responsible for preventing the collapse of the galactic disk. Both simulated galaxies show similar vertical ram and thermal pressures at all three radii. The thermal pressure has a double peak or a flat top at the galactic midplane whose origin we attribute to the multiphase nature and multi-temperature structure of the dense ISM in our simulation. The magnetic stress on the other hand only shows a single central peak. The midplane magnetic stress is larger in the MHD simulation in comparison to the CRMHD simulation in the galactic centre region and the outer disk. At intermediate radii in the galaxy, the CRMHD simulation reaches slightly higher midplane magnetic field stresses. At all galactic radii, the median magnetic stress is larger than the median thermal pressures and our simulated ISM is hence in a regime of low plasma-$\beta$ values, where $\beta = P_\mathrm{th} / P_\mathrm{mag}$ and $P_\mathrm{mag}=B^2/8\pi$. Whether the magnetic fields significantly contribute to stopping the gravitational collapse warrants its own detailed investigation. Works by \citet{2018GirichidisII} and \citet{2023Whitworth} show that magnetic fields support against collapse perpendicular to the magnetic field direction in dense gas ($n_\mathrm{H} \gtrsim 1~\mathrm{cm}^{-3}$). The CR pressure is larger than all other pressure components except for the galactic main body where the magnetic stress is still the dominant component. However, the scale height of the CR pressure profile is enlarged inside the galactic disk if compared to the other pressure and stress profiles which share similar scale heights.

Pressure and stresses in the inner CGM show more complicated profiles. The magnetic stress drops following its ISM profile and has negligible values in the inner CGM, while the other components radically increase their scale height above $~2$~kpc and prevail even above the galactic disk. A similar behavior for the magnetic stress is observable in higher resolution tallbox simulations \citep{2020Vijayan, 2022Armillotta}. The fluctuations in the CRMHD simulation of the magnetic stress hint towards a more effective entrainment of magnetic field lines in comparison to the MHD simulation. For the galaxy in the CRMHD simulation, the CR pressure is the strongest pressure component at all galactic radii. \citet{2021Armillotta} shows that the vertical distribution of CRs varies strongly with the modelled CR transport mechanisms. In our simulation, we only consider our full transport model but get overall different shapes of the CR pressure profile at different galactic radii.

Around the apparent phase transition, multiple profiles show dips. We already identified that this point marks the heights to which galactic fountain flows are present in our simulation and are mixed with the dilute outflow. Both phases contribute to the underlying statistics below this galactic height but only the dilute outflow influences the pressure statistics above this point. This causes the transition at around $2-2.5$~kpc. The thermal pressure is generally higher in the MHD simulation if compared to the CRMHD counterpart. This is in line with the temperature and density profiles in Fig.~\ref{fig:density_temperaure_profile} which show that the density drops slower than the temperature increases around this point.

Another important feature of the pressure profiles is visible at galactic heights $5-7.5$~kpc where the ram pressure starts to increase again while the thermal pressure starts to decrease more rapidly. Identifying this location in the outflow velocity profiles of Fig.~\ref{fig:force_balance} reveals that at these galactic heights, the galactic winds start to show strong vertical gradients in the vertical velocity $\varv_z$. This explains the increase in ram pressure but also the decrease in thermal support which is caused by adiabatic expansion. 

\section{Kinematic phase structure}
\label{sec:phase_structure}

\begin{figure*}
   \centering
   \resizebox{\hsize}{!} { \includegraphics[width=\textwidth]{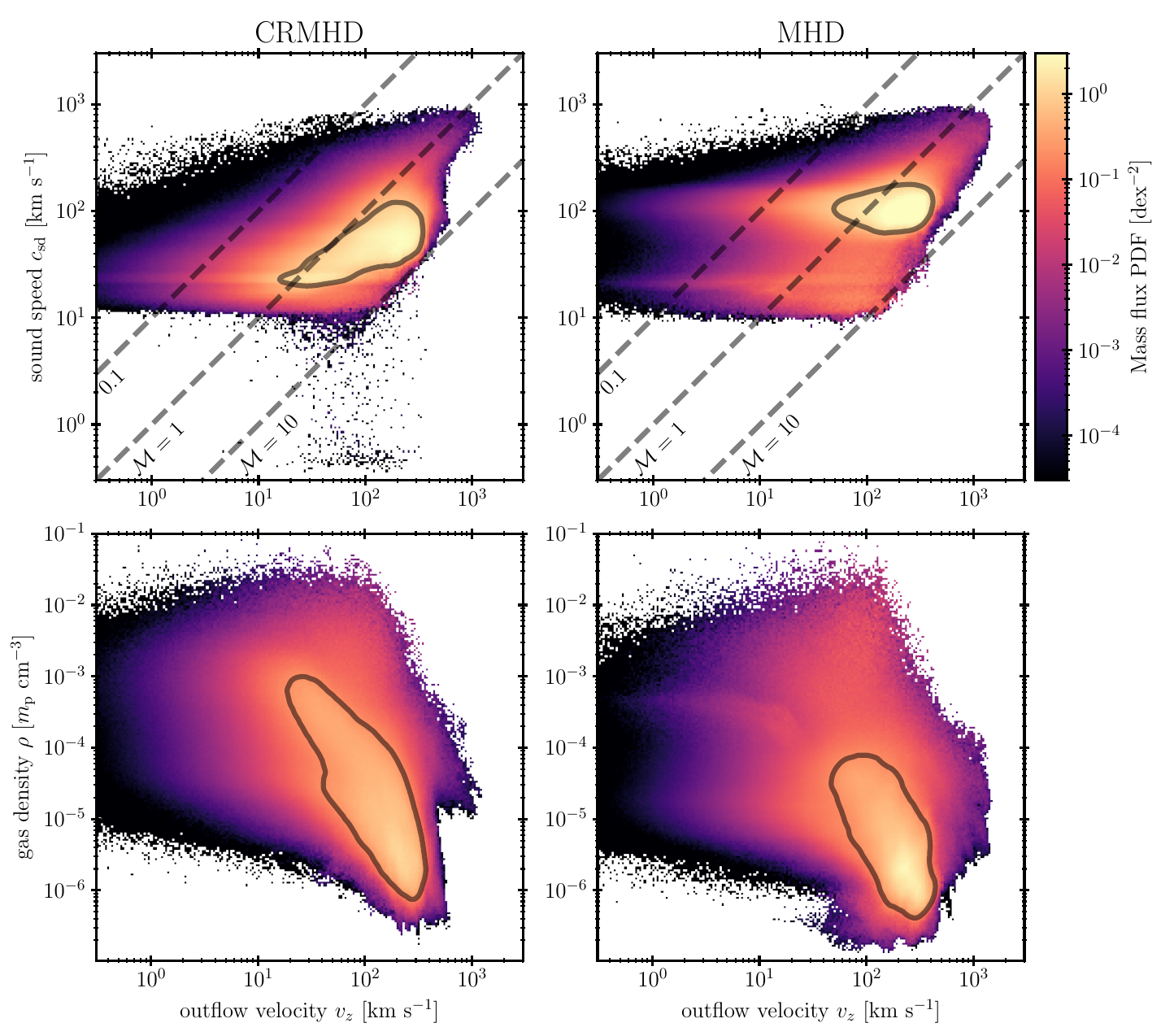} }
   \caption{Phase structure of the galactic outflows in both the CRMHD (left) and MHD (right) simulations. We show the probability distribution function of the volume-weighted mass flux $\rho \varv_z$ to highlight gas that is responsible for the mass loss from the galaxy. The top panels show the sound speed (which can also be seen as a proxy for the gas temperature) of gas inside $R < 30$~kpc and $3~\mathrm{kpc} < |z| < 100\mathrm{kpc}$ while the bottom panels show the hydrogen number density over outflow velocity $\varv_z$. We mark the Mach numbers 0.1, 1, and 10 which can be used to infer that the outflow in both simulations is supersonic. The contour in each panel encloses 75\% of the total mass flux. The phase structure of the CRMHD is more diverse and spans multiple orders of magnitude in density while being cooler than the MHD simulation.}
   \label{fig:phase_structure}
\end{figure*}

In this section, we investigate the difference between the two simulations in their resulting phase structure of the galactic outflows. In Fig.~\ref{fig:phase_structure}, we plot the two-dimensional probability distribution function (PDF) of the volume-weighted mass flux $\rho \varv_z$ in the sound-speed $c_\mathrm{sd}$ and outflow velocity $\varv_z$ plane as well as the mass density $\rho$ and outflow velocity $\varv_z$ plane. This allows us to identify the thermodynamic properties of the mass-carrying components of the galactic winds. In our analysis, we only account for gas that is above the galactic disk ($R < 30$~kpc) that is weakly influenced by the near-plane fountain flows ($3~\mathrm{kpc} < |z| < 100~\mathrm{kpc}$). We also show a contour in each panel which contains 75\% of the total mass flux.

The galactic winds in both simulations are supersonic with Mach numbers ranging from 0.1 to 10 while most of the mass flux is contained between Mach numbers of 1 to 10. The gas inside the CR-driven galactic wind reaches higher Mach numbers in comparison to the gas in the MHD simulation. As both winds have roughly the same outflow velocity profiles, see Fig.~\ref{fig:force_balance}, this difference is mainly caused by the difference in temperature in both winds. The majority of the mass flux in the MHD simulation has sound speeds $c_\mathrm{sd} \sim 100$~km~s$^{-1}$, corresponding to temperatures of $10^6$~K following the temperature profile displayed in Fig.~\ref{fig:density_temperaure_profile}. In the CR-driven galactic wind, this is distributed over a broader range of sound speeds $10~\mathrm{km}~\mathrm{s}^{-1} < c_\mathrm{sd} < 100~\mathrm{km}~\mathrm{s}^{-1}$. This indicates that the outflow in the CRMHD is multiphase in nature.

Compared to higher-resolution simulation \citep[e.g.,][]{2023Rathjen,2024Steinwandel}, we are lacking gas in the warm phase with temperatures of around $10^4$~K (or equivalently $c_\mathrm{sd} \sim 10$~km~s$^{-1}$). This indicates insufficient resolution to maintain a warm and hot phase co-spatially and to preclude artificial mixing of these phases, which occurs in our simulation.
In the CRMHD simulation, there are data points at low temperatures $c_\mathrm{sd} \sim 1$~km~s$^{-1}$ and high Mach numbers $\mathcal{M} > 10$ which are too sparse to draw conclusions from them as they might be numerical noise. Further and higher resolution simulations of the inner-CGM hosting CR-driven galactic winds are needed to explore the full thermodynamic nature of this medium.

In the bottom panels of Fig.~\ref{fig:phase_structure}, we display the kinematic statistics of the number density using the mass flux PDF as before. Densities in the outflow span multiple orders of magnitude ranging from $10^{-2}$ to $10^{-7}$ particles per cm$^3$. Gas with lower densities tends to be outflowing faster. This is consistent with the picture of a continuously accelerated wind that expands into the CGM. The majority of mass loss in both simulations is carried by low-density gas with $\rho \sim 10^{-6}$ to $10^{-5}~m_\mathrm{p}~\mathrm{cm}^{-3}$. Nevertheless, the distribution of mass-carrying densities encloses three orders of magnitude for the CRMHD simulation while these densities only span two orders of magnitude in the MHD simulation. This increase is mostly caused by denser gas in the CRMHD run which can be entrained in the CR-supported galactic wind. This again indicates that the outflow above the CRMHD galaxies is more diverse and richer in dynamics.

\begin{figure*}
   \centering
   \resizebox{\hsize}{!} { \includegraphics[width=\textwidth]{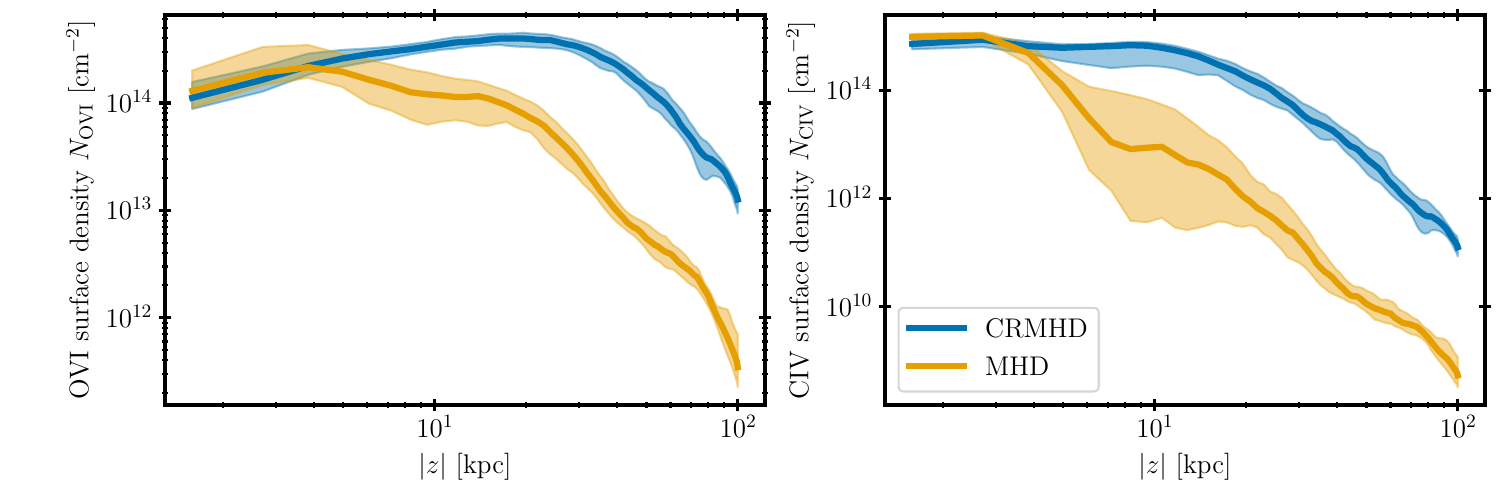} }
   \caption{Surface density profiles of the \ion{O}{vi} and \ion{C}{iv} ions in the galactic outflow ($R < 30$~kpc) for the CRMHD (left) and MHD (right) simulations. We show the median of the surface densities as thick lines and their 20$^\mathrm{th}$ to 80$^\mathrm{th}$ percentiles using shaded areas. Both metal ions trace the warm-hot component ($T\sim 10^{5} - 10^{6}$~K) of the CGM. The different phase structures of both galactic winds lead to differences in the abundance of both ions. }
   \label{fig:metal_abundances_profile}
\end{figure*}

\section{Metal absorption lines}
\label{sec:metal_absorption}

The different phase structures of the galactic winds should be discernible using metal line absorption along sight-lines through the CGM. Various metal ions can be used to detect and infer the physical state of this medium using their absorption of radiation originating from background sources \citep{2017Tumlinson}. Observational studies, such as AMIGA \citep{2020Lehner} or COS-HALOES \citep{2014Werk}, collecting multiple sight-lines map the CGM using these metal absorption components. We use this section to compare both of our simulations in terms of their absorption properties of \ion{O}{vi} and \ion{C}{iv} ions which are highly ionized and probe the temperature regime of the galactic winds. 

To this end, we calculate the ionization equilibrium of carbon and oxygen in post-processing and account for radiative recombination \citep[rates taken from][]{2006Badnell} and dielectronic recombination \citep[rates taken from][]{2003Badnell} which is balanced by collisional ionization by free electrons \citep[rates taken from][]{1997Voronov} and photoionization \citep[cross-sections taken from][]{1996Verner} caused by the meta-galactic UV background \citep[specific energy density taken from][]{2019Puchwein}. This time-independent approach underestimates the abundance of these ions at the lower temperature end of our parameter space, at $T\sim10^5$~K, due to the absence of the recombination lag in our model \citep{2007Gnat, 2011Vasiliev}. 

In Fig.~\ref{fig:metal_abundances_profile}, we show projected surface densities as a function of height of the galactic outflow regions in both simulations, averaged across a cylindrical radius $R < 30$~kpc and over 10 snapshots from 90 to 100 Myr. We display the median as well as the 20$^\mathrm{th}$ to 80$^\mathrm{th}$ percentiles. At the interface of ISM and CGM in between 1 and 3 kpc, both simulations show the same amount of projected \ion{O}{vi} and \ion{C}{iv}. These ions are mainly contained inside high-temperature SN bubbles that are currently rising from the galactic disk. Their dynamics are thus regulated by the internal dynamics of the ISM. As both galaxies have similar masses, morphologies, and SFRs, this similarity between both simulations is expected. At larger galactic heights, above the apparent phase transition point and inside the galactic wind, both simulations show different trends. The galactic wind in the CRMHD simulation entrains more \ion{O}{vi} and \ion{C}{iv} compared to the galactic wind in the MHD simulation. This difference cannot solely be explained by the increased gas density in the CRMHD outflow, cf.\ Fig.~\ref{fig:density_temperaure_profile}, which amounts to $\sim0.5$ dex. The maximum difference of $\sim 1.5$ dex between both galactic winds is reached at $\sim50$~kpc for the \ion{O}{vi} ion. The increased surface density of these ions in the CRMHD simulation is thus caused by a combination of increased density and lower temperature. In conclusion, the inclusion of CRs in the wind driving and the resulting difference in the multiphase phase structure of the galactic outflow have observable consequences. 

Similar findings are reported in 
\citet{2016Salem}, who find that the COS-HALOES \ion{O}{vi} column densities are better matched by a cosmological simulation that incorporates CR feedback, which stabilizes the CGM and allows it to transition to a cooler and denser state in comparison to the CGM in a similar simulation without CRs. Other simulations projects also report an increased amount of \ion{O}{vi} in galactic haloes that are affected by CR feedback \citep{2020Ji, 2024DeFelippis}. Furthermore, the kinematic structure of CGM absorption lines is better matched in CR pressure-supported halos \citep{2021Butsky}. Simulations including only thermally-mediated feedback predict too little \ion{O}{vi} in galactic halos \citep{2016Oppenheimer, 2016Corlies} or \ion{O}{vi} absorption parameters consistent with observations \citep{2018Nelson, 2022Damle}.

\section{Summary and conclusion}

In this paper, we investigate the difference between thermally- and CR-driven galactic winds by comparing a pair of simulations of a Milky Way-type isolated galaxy, one accounting for CR feedback while the other does not. The simulations are conducted with the \textsc{Arepo} moving-mesh code and use the newly developed \textsc{CRISP} feedback framework which models the relevant microphysical chemical and thermodynamical processes inside the ISM that are responsible for shaping this medium. These processes also determine how CRs are transported within and outside a galaxy. We employ a super-Lagrangian refinement scheme inside the galactic outflows which tremendously increases the numerical resolution in this medium and allows for a detailed statistical evaluation of outflow quantities such as mass and energy loading factors, the relevant gas and CR pressures, as well as the acting forces which ultimately drive the winds. The findings of our detailed analysis are as follows:
\begin{enumerate}
\item Irrespective of whether we account for CR feedback or not, our simulated galaxies always drive a galactic wind. The galactic outflow in the MHD simulation is entirely driven by a thermal pressure gradient powered by SNe while the galactic outflow in the CRMHD case is enhanced due to the additional forces provided by the CRs. The outflows drastically differ in their morphological appearance. The outflow in the CRMHD simulation is multiphase and highly structured with a colder cloudlet population embedded into dilute gas at intermediate temperatures ($\sim 10^5$~K). In result, the galactic wind of the CRMHD simulation entrains more gas to larger galactic heights (see Fig.~\ref{fig:gallery_3}). By contrast, the MHD galaxy launches a high temperature and underdense galactic outflow consisting of multiple plume-like structures (see Figs.~\ref{fig:gallery} and \ref{fig:density_temperaure_profile}). 

\item Interestingly, the additional feedback provided by CRs has little impact on the gas morphology of the ISM which is mostly shaped by the effective mechanical impact of SN bubbles on their surrounding diffuse or cold filamentary ISM (see Fig.~\ref{fig:gallery_2}). This observation is supported by the similarity of the SFR in both simulations, which is rather unaffected by the presence of CRs inside the galaxy due to their large pressure gradients and small fluxes. This is in line with cosmological galaxy formation simulations using the \textsc{Auriga} feedback model \citep{2020Buck}.

\item Because of the similar SFR in both simulations, a similar amount of free energy is injected into the ISM surrounding SNe. Hence, both galaxies should have the same potential to drive a mass-loaded galactic wind but the mass-loading of the CR-driven galactic wind is enhanced by a factor of $\sim2$. Likewise, more energy is entrained in the CRMHD galactic wind, mostly contained in the form of CRs. Despite the increased mass-loading of $\eta_{M}\sim0.5$, the SFRs of both galaxies decline at the same rate. This indicates that star formation is the dominant mechanism that depletes gas inside the ISM and not galactic outflows (see Fig.~\ref{fig:loading_factors}).

\item The released amount of mass and energy of a galactic outflow builds up a pressurized inner CGM with almost flat thermal and ram pressure profiles between galactic heights of $2.5 - 10$~kpc. Including CRs as a feedback channel leaves the internal pressure structure of the galactic disk almost unaltered. Vertical magnetic stresses, thermal and ram pressures show steeply declining profiles inside the ISM, which are able to counteract the gravity exerted by the disk. The resulting CR pressure profile is smooth and only slowly declines from the ISM into the CGM where it is the dominant pressure component (see Fig.~\ref{fig:pressure_profile}).

\item Collective SN feedback stirs turbulence inside the ISM and launches fountain flows in both galaxies, which expel dense gas to galactic heights of approximately $2 - 3$~kpc with excursions up to 10 kpc. This leads to an apparent phase transition of the outflowing material but merely marks the point above which no cold ($T\sim10^4$~K) gas can be found in our simulations at our resolution (see Fig.~\ref{fig:density_temperaure_profile}).

\item The galactic winds are launched slightly above the point of the apparent phase transition. In the MHD simulation, the thermal pressure gradient is able to overcome gravity while in the CRMHD galaxy the combined action of the thermal and CR pressure gradients are contributing to the acceleration of the galactic winds. The outflows show similar outflow velocity profiles reaching values between $200 - 300 ~\mathrm{km}~\mathrm{s}^{-1}$ above 50 kpc (see Fig.~\ref{fig:force_balance}).

\item The phase structure of both winds in quite different. The multiphase wind of the CRMHD galaxy entrains gas with hydrogen densities ranging from $10^{-6}$ to $10^{-3}~\mathrm{cm}^{-3}$ and temperatures from $10^4$ to $10^6$~K. The outflow from the MHD galaxy is more uniform and characterized by a single hot phase with temperatures of $\sim10^{6}$~K and hydrogen number densities between $10^{-6}$ to $10^{-5}~\mathrm{cm}^{-3}$ (see Fig.~\ref{fig:phase_structure}).

\item The difference in temperature and density of the galactic winds has a direct consequence for their observables. We examine the absorption properties of \ion{O}{vi} and \ion{C}{iv} by analysing their surface densities inside the galactic wind and find that CR-driven galactic winds are able to carry ten times more \ion{O}{vi} and \ion{C}{iv} in the observable range of $\Sigma_\mathrm{\ion{O}{vi}}$ and $\Sigma_\mathrm{\ion{C}{iv}}$ from $10^{14}$ to $10^{15}~\mathrm{cm}^{-2}$ (see Fig.~\ref{fig:metal_abundances_profile}).

\end{enumerate}

In summary, we find that CRs accelerated at SNe inside the ISM leave the galaxy to build a global CR pressure distribution around the galactic disk which supports a dense multiphase galactic outflow. Sole SN-mediated kinetic and thermal feedback stirs turbulence inside the galactic disk and launches galactic fountains flows but expels less mass and energy into the galactic outflow.

\begin{acknowledgements}
TT and CP acknowledge support by the European Research Council under ERC-AdG grant PICOGAL-101019746. This work was supported by the North-German Supercomputing Alliance (HLRN) under project bbp00046 and bbp00070.
\end{acknowledgements}

%
%

\bibliographystyle{aa}
\bibliography{main}

\end{document}